\documentclass[usenatbib,usegraphicx]{mn2e}
\usepackage[totalwidth=480pt, totalheight=680pt]{geometry}

\title[A high-frequency radio survey of LLAGNs]{A high-frequency radio survey of low-luminosity active galactic nuclei}
\author[A. Doi, S. Kameno, K. Kohno, K. Nakanishi, and M. Inoue]{A. Doi$^{1,2}$\thanks{E-mail:doi@yamaguchi-u.ac.jp}\thanks{Present address: Department of Physics, Faculty of Science, Yamaguchi University, Yoshida, Yamaguchi 753-8512, Japan.}, S. Kameno$^{2}$, K. Kohno$^{3}$, K. Nakanishi$^{4}$, and M. Inoue$^{2}$\\
$^{1}$Department of Astronomy, University of Tokyo, 7-3-1 Hongo, Bunkyo-ku, Tokyo 113-0033, Japan\\
$^{2}$National Astronomical Observatory, 2-21-1 Osawa, Mitaka, Tokyo 181-8588, Japan\\
$^{3}$Institute of Astronomy, The University of Tokyo, 2-21-1 Osawa, Mitaka, Tokyo 181-0015, Japan\\
$^{4}$Nobeyama Radio Observatory, Minamimaki, Minamisaku, Nagano 384-1305, Japan}

\begin{document}

\date{Accepted for publication in MNRAS}

\pagerange{\pageref{firstpage}--\pageref{firstpage}} \pubyear{2005}

\maketitle

\label{firstpage}

\begin{abstract}
We investigate the high-frequency radio spectra of twenty low-luminosity active galactic nuclei (LLAGNs) with compact radio cores.  Our millimetre survey with the Nobeyama Millimetre Array (NMA) and analyses of submillimetre archival data that had been obtained with the Submillimetre Common User Bolometer Array (SCUBA) on the James Clerk Maxwell Telescope (JCMT) reveal the following properties.  At least half of the LLAGNs show inverted spectra between 15 and 96~GHz; we use published data at 15~GHz with the Very Large Array (VLA) in a 0.15-arcsec resolution and our measurements at 96 GHz with the NMA in a 7-arcsec resolution.  The inverted spectra are not artificially made due to their unmatched beam sizes, because of little diffuse contamination from dust, \mbox{H\,{\sc ii}} regions, or extended jets in these LLAGNs.  Such high-frequency inverted spectra are apparently consistent with a `submillimetre bump', which is predicted by an advection-dominated accretion flow (ADAF) model.  We find a strong correlation between the high-frequency spectral index and low-frequency core power measured with very-long-baseline-interferometry (VLBI) instruments.  The inverted spectra were found exclusively in low-core-power sources, while steep spectra were in high-core-power ones with prominent pc-scale jets.  This suggests that the ADAF and nonthermal jets may coexists.  The flux ratios between disc and jet seem to be different from LLAGN to LLAGN; disc components can be seen in nuclear radio spectra only if the jets are faint.
\end{abstract}


\section{INTRODUCTION}
Low-luminosity active galactic nuclei (LLAGNs), operationally defined as a H$\alpha$ luminosity of $< 10^{40}\ \mathrm{ergs\ s^{-1}}$, reside in about 40~per~cent of nearby bright galaxies, and occupy majority of the population of AGNs \citep{Ho_etal.1997a,Ho_etal.1997b}.  Its low luminosities are thought to be caused by very low accretion rate on to super-massive black holes.  Such accretion has often been explained by the model of optically-thin disc such as an advection-dominated accretion flow (ADAF; \citealt{Ichimaru1977,Narayan_Yi1994,Abramowicz_etal.1995}), rather than optically-thick disc (`standard disc'; \citealt{Shakura&Sunyaev1973}).  It is predicted that the ADAF has a `submillimetre bump' in its spectral-energy distribution (SED), while the standard disc is responsible for a big-blue bump in the SED of luminous AGNs such as quasars.  Since a high brightness temperature of $\sim10^{9.5}$~K in radio bands is expected from the ADAF model, the first imaging investigations for accretion discs must be promising with future very-long-baseline-interferometry (VLBI) instruments that will provide micro-arcsec resolutions \citep{Hirabayashi_etal.2002,Ulvestad2000,Miyoshi_etal.2003}.  Therefore, the observational evidence of the ADAF is crucial in the LLAGN radio sources.

Although the ADAF model successfully explains the broadband spectra of LLAGNs, there is a significant gap between observations and the ADAF spectrum especially at low-frequency radio bands (e.g., \citealt{Manmoto_etal.1997,Quataert_etal.1999,Ho1999}).  This indicates that additional unknown emission does exist, and putative emission from the accretion disc may be buried under it.  The submillimetre bump means a highly inverted spectrum at centimetre-to-submillimetre wavelengths of spectral index $\alpha\sim0.4$, where $S_\nu \propto \nu^{+\alpha}$, $S_\nu$ is flux density at frequency $\nu$ \citep{Mahadevan1997}.  Observations for LLAGNs have been carried out exclusively at centimetre bands where high sensitivities are available, because most of LLAGNs are very faint radio sources.  About half of low-luminosity Seyfert galaxies and low-ionization nuclear emission-line regions (LINERs; \citealt{Heckman1980}) hold a compact radio core \citep{Nagar_etal.2000,Nagar_etal.2002}, and at least 25 per cent of transition objects, which are hypothesized to be LINER/\mbox{H\,{\sc ii}}-nucleus composite systems \citep{Ho_etal.2003,Ho1996}, also hold a compact core \citep{Filho_etal.2000,Filho_etal.2002}, at 15 or 8.4~GHz in 0.15--2.5 arcsec resolutions.  Most of the cores show nearly flat spectra ($\alpha \sim 0$) in 0.5-arcsec resolution: that is evidence for jet domination \citep{Nagar_etal.2001}.  Although slightly inverted spectra have been found in unresolved compact cores of several LLAGNs in milli-arcsec resolutions of VLBIs, their radio-to-X-ray luminosity ratios suggest that another significant radio component seems to contribute in addition to original fluxes from the ADAF model \citep{Ulvestad_Ho2001a,Anderson_etal.2004}.  To explain this radio excess, an jet--ADAF model has been proposed \citep{Falcke_Biermann1999,Yuan_etal.2002}.  Thus, at centimetre bands, the contamination that is believed to be from jets has prevented the ADAF from being revealed.

In the present paper, we report millimetre survey and submillimetre data analyses for many LLAGNs.  Although technically difficult, a high-frequency observation is very promising to detect the inverted spectrum of the ADAF because of following two advantages: (1) spectral measurements at high frequencies are less affected by non-thermal jets, which generally show power-law spectra ($-\alpha=0.75$--$0.5$), and (2) the ADAF predicts larger flux densities at higher frequencies.  In fact, it has been obsevationally confirmed that flux densities at millimetre wavelengths are 5--10 times larger than those at centimetre wavelengths in Sgr~A* (e.g. \citealt{Falcke_etal.1998}), which is the nearest low-accretion rate system from us.  However, contamination from diffuse dust may be harmful at $\ga100$~GHz when we use a poor spatial resolution.  Therefore, the use of a beam as small as possible and the estimations of every conceivable contamination are essential.  

The present paper is structured as follows.  Sample selection is described in Section~\ref{section:sample}.  Observations and data reduction are described in Section~\ref{section:obs&reduction}.  In Section~\ref{section:Results}, we report these results and the estimations of diffuse contamination in our millimetre measurements.  Our results of submillimetre photometry are utilized only for estimation for dust contribution.  The origins of the spectra are discussed in Section~\ref{section:discussion} from the relation between the high-frequency radio spectrum and low-frequency radio core power.  Finally, we summarize in Section~\ref{section:summary}.

\section{Sample}\label{section:sample}
Our sample, `VLBI-detected LLAGN sample', covers 20 out of 25 all-known LLAGN radio sources that have been detected with VLBIs (column [1] of Table~\ref{table1}).  The other five LLAGNs are NGC~4235, NGC~4450 \citep{Anderson_etal.2004}, NGC~524, NGC~5354, and NGC~5846 \citep{Filho_etal.2004}, which were recently newly VLBI-detected, but whose reports had not yet been published when we plan the survey.  It is difficult to estimate resultant selection bias including into our sample, because of multiplicity in selection criteria across the past VLBI surveys.  However, at least we can say that 16 out of 20 sources of our sample are all known LLAGNs that are at $<20$~Mpc with $>2.7$~mJy at 15~GHz, and the other four sources (NGC 266, NGC 3147, NGC 3226, and NGC 4772) are more distant.  It may be, therefore, as a moderately good distance-limited and radio-flux-limited sample for LLAGNs with a compact radio core.

\section{Observations and data reduction}\label{section:obs&reduction}

\subsection{Millimetre band}\label{subsection:3mm}
We made millimetre continuum observations for 16 out of 20 targets using the Nobeyama Millimetre Array (NMA), at the Nobeyama Radio Observatory (NRO), with D configuration that is the most compact array configuration.  We had precluded the other four sources from our observation list because they had already been observed at $\sim3$~mm in the past (NGC~266, \citealt{Doi_etal.2005}; NGC~3031, \citealt{Sakamoto_etal.2001}; NGC~4258, Doi~et~al. in prep.; NGC~4486, \citealt{Despringre_etal.1996}).  Our campaign spent more than 20 observational days between 2002 November~28 and 2003 May~25.  Most of weak targets were observed over several days for integration.  Visibility data were obtained with double-sided-band receiving system at centre frequencies of 89.725 and 101.725 GHz, which were Doppler-tracked.  We used the Ultra Wide-Band Correlator \citep[UWBC;][]{Okumura_etal.2000}, which can process a bandwidth of 1~GHz per each sideband, i.e., 2~GHz in total.  The wide bandwidth gives the NMA a very high sensitivity for continuum observations.  Even if the systemic velocities of our targets and galaxies' rotations of several hundred km~s$^{-1}$ are accounted for, the observing band can avoid possible contaminations from several significant line emissions: $^{13}$CO($J=1-0$), C$^{18}$O($J=1-0$), HCN($J=1-0$), HCO$^{+}$($J=1-0$), and SiO($J=2-1$).  A system noise temperature, $T_\mathrm{sys}$, was typically about 150~K in the double-sided bands.  For observations of each target, we made scans of a reference calibrator close to the target every 20 or 25~minutes for gain calibration.  Bandpass calibrators of very bright quasars were scanned once a day.

The data were reduced using the UVPROC-II package \citep{Tsutsumi_etal.1997}, developed at the NRO, by standard manners, including flagging bad data, baseline correction, opacity correction, bandpass calibration and gain calibration.  To achieve higher sensitivity, visibilities of both the sidebands were combined in the same weight, which resulted in a centre frequency of 95.725 GHz.  Each daily image was individually made in natural weighting and deconvolved using the \textsc{AIPS} software, developed at the National Radio Astronomy Observatory (NRAO).  Statistically significant variability was detected from the daily images in several LLAGNs, which will be reported in a separate paper.  In the present paper, the visibilities of all days were combined by weighting with $T_\mathrm{sys}^{-2}$, and then imaged.  The half-power beam widths (HPBWs) of synthesized beams were typically $\sim7$~arcsec, corresponding to $\sim$660~pc at a mean distance of 18.9~Mpc in our sample.  The flux scales of the gain calibrators were derived with the uncertainty to 10~per~cent by relative comparisons to the flux-known calibrators, such as Uranus, Neptune, or Mars.  These were scanned quasi-simultaneously when they were at almost the same elevations.  The rms of noise on the images were estimated from statistics on off-source blank sky with the \textsc{IMEAN} task of the \textsc{AIPS}.  Source identifications and measurements of flux densities were done on image domain by elliptical Gaussian profile fitting with the \textsc{JMFIT} task of the \textsc{AIPS}.  We derived total errors in the flux measurements from root sum square of the errors in the Gaussian fitting (including thermal noise) and the flux scaling.

\subsection{Submillimetre band}\label{subsection:850um}
We attempted to measure flux densities at submillimetre band from the nuclear regions using archival data.  We searched the data that had been obtained at 347.38 GHz ($850\ \mathrm{\mu m}$) using Submillimetre Common User Bolometer Array (SCUBA; \citealt{Holland_etal.1999}) on the James Clerk Maxwell Telescope (JCMT).  The data of 14 out of 20 objects of the VLBI-detected sample were available to the public on 2004.  The JCMT measurements for five objects have already been reported (NGC 266, \citealt{Doi_etal.2005}; NGC 4258, Doi et~al. in~prep.; NGC 4374, \citealt{Leeuw_etal.2000}; NGC 4472, \citealt{diMatteo_etal.1999}; NGC 4486, \citealt{Knapp_Patten1991}).

Four objects had been observed in jiggle-map mode.  We reduced these data using the SCUBA User Reduction Facility (\textsc{SURF}) package.  Standard reduction procedures were used, including flat fielding, flagging of transient spikes, correction for extinction, pointing correction, sky removal and flux-density scaling.  Flux calibrators (Uranus, Mars, or CRL~618) gave us flux scaling factors (we assumed its uncertainty to $\sim$15~per~cent) and an effective beam size of 15.1~arcsec.  The derivations of the rms of noise on images, source identifications, measurements of flux densities, and total error estimations were done in the same manners as the NMA reductions described in Section~\ref{subsection:3mm}.

The other five objects had been observed in photometry mode.  We followed standard reduction procedures, including flat fielding, flagging of transient spikes, correction for extinction, sky removal, averaging, and flux-density scaling, using the \textsc{SURF} package.  The photometry can measure a flux density from a beam-size region at the nucleus, i.e., an intensity.  We derived total errors in flux measurements from root sum square of the errors of the image noise and the flux scale.

\section{Results}\label{section:Results}

\begin{table*}
\defcitealias{Doi_etal.2005}{a}
\defcitealias{Sakamoto_etal.2001}{c}
\defcitealias{Despringre_etal.1996}{e}
\defcitealias{Leeuw_etal.2000}{f}
\defcitealias{diMatteo_etal.1999}{g}
\defcitealias{Knapp_Patten1991}{h}

\begin{center}
\begin{minipage}{160mm}
\caption{Observation results for the VLBI-detected LLAGN sample.}

\begin{tabular}{lrccrrccc} 
\hline
\multicolumn{1}{c}{Name} & \multicolumn{1}{c}{$S_\mathrm{96GHz}$} & $\sigma^\mathrm{rms}_\mathrm{96GHz}$ & Ref. & \multicolumn{1}{c}{$I^\mathrm{peak}_\mathrm{347GHz}$} & \multicolumn{1}{c}{$S_\mathrm{347GHz}$} & $\sigma^\mathrm{rms}_\mathrm{347GHz}$ & Mode & Ref. \\
\multicolumn{1}{c}{} & \multicolumn{1}{c}{(mJy)} & (mJy beam$^{-1}$) &  & \multicolumn{1}{c}{(mJy beam$^{-1}$)} & \multicolumn{1}{c}{(mJy)} & (mJy beam$^{-1}$) &  &  \\
\multicolumn{1}{c}{(1)} & \multicolumn{1}{c}{(2)} & (3) & (4) & \multicolumn{1}{c}{(5)} & \multicolumn{1}{c}{(6)} & (7) & (8) & (9) \\\hline
NGC 266 & $<2.7$ & 0.9  & \citetalias{Doi_etal.2005} & $<18.5$ & $<18.5$ & 6.2 & M & a \\
NGC 2787 & $15.4\pm2.4$ & 1.0  & b & $34.3 \pm 5.9$ & \ldots & 3.0 & P & b \\
NGC 3031 & $390\pm39^{*1}$ & \ldots & \citetalias{Sakamoto_etal.2001} & $86.6 \pm 13.2$ & \ldots & 2.1 & P & b \\
NGC 3147 & $3.4\pm1.5$ & 0.9  & b & $50.7 \pm 9.3$ & $1747 \pm 178$ & 5.4 & M & b \\
NGC 3169 & $8.1\pm4.0$ & 1.7  & b &  &  &  &  &  \\
NGC 3226 & $8.4\pm1.8$ & 1.0  & b &  &  &  &  &  \\
NGC 3718 & $11.1\pm3.5$ & 1.9  & b & $19.1 \pm 3.8$ & \ldots & 2.6 & P & b \\
NGC 4143 & $6.7\pm2.7$ & 1.2  & b &  &  &  &  &  \\
NGC 4168 & $6.0\pm2.5$ & 1.3  & b &  &  &  &  &  \\
NGC 4203 & $15.3\pm2.9$ & 1.0  & b & $<10.24$ & \ldots & 3.4 & P & b \\
NGC 4258 & $11.0\pm2.7^{*2}$ & \ldots & d & $93.7 \pm 29.3$ & $3377 \pm 254$ & 10.5 & M & d \\
NGC 4278 & $57.0\pm8.5$ & 3.1  & b & $49.4 \pm 7.4$ & $53.3 \pm 8.0$ & 4.4 & M & b \\
NGC 4374 & $147\pm17$ & 5.0  & b & $144 \pm 13^{*4}$ & \ldots &  & PM & \citetalias{Leeuw_etal.2000} \\
NGC 4472 & $15.6\pm5.4$ & 2.1  & b & $3.0 \pm 1.3$ & \ldots &  & P & \citetalias{diMatteo_etal.1999} \\
NGC 4486 & $1880\pm100^{*3}$ & \ldots & \citetalias{Despringre_etal.1996} & $1069 \pm 39$ & \ldots &  & P & \citetalias{Knapp_Patten1991} \\
NGC 4552 & $12.2\pm4.1$ & 2.6  & b & $<26.6$ & $<26.6$ & 8.9 & P & b \\
NGC 4565 & $<4.7$ & 1.6  & b &  &  &  &  &  \\
NGC 4579 & $28.6\pm8.2$ & 2.7  & b & $85.7 \pm 23.8$ & $150 \pm 30$ & 7.5 & M & b \\
NGC 4772 & $7.1\pm2.4$ & 1.1  & b &  &  &  &  &  \\
NGC 5866 & $<2.9$ & 1.0  & b & $51.7 \pm 19.1$ & $173 \pm 31$ & 3.7 & M & b \\\hline
\end{tabular}
\label{table1}

\textbf{*1} --- mean flux density of multi-epoch data at 103 GHz.\\
\textbf{*2} --- mean flux density of multi-epoch data at 101 GHz.\\
\textbf{*3} --- obtained with the IRAM Plateau de Bure interferometer at 89 GHz.\\
\textbf{*4} --- mean flux density of multi-epoch data.

\smallskip
Note. --- Column are: \textbf{(1)} galaxy name; \textbf{(2)} flux density at 96 GHz; \textbf{(3)} r.m.s.~of image noise at 96 GHz; \textbf{(4)} reference for flux density at 96 GHz; \textbf{(5)} peak intensity of nuclear component at 347 GHz; \textbf{(6)} integrated flux density at 347 GHz; \textbf{(7)} r.m.s.~of image noise at 347 GHz; \textbf{(8)} observation mode of the SCUBA at 347 GHz, `M' represents jiggle mapping, `P' represents photometry; \textbf{(9)} reference for data at 347 GHz.

\smallskip
Note. --- Reference:
\textbf{\citetalias{Doi_etal.2005}}: \citealt{Doi_etal.2005};
\textbf{b}: the present paper;
\textbf{\citetalias{Sakamoto_etal.2001}}: \citealt{Sakamoto_etal.2001};
\textbf{d}: Doi et~al. in prep.;
\textbf{\citetalias{Despringre_etal.1996}}: \citealt{Despringre_etal.1996};
\textbf{\citetalias{Leeuw_etal.2000}}: \citealt{Leeuw_etal.2000};
\textbf{\citetalias{diMatteo_etal.1999}}: \citealt{diMatteo_etal.1999};
\textbf{\citetalias{Knapp_Patten1991}}: \citealt{Knapp_Patten1991}
\end{minipage}
\end{center}
\end{table*}

Results of the NMA observations and the analyses of JCMT archival data are listed in Table~\ref{table1}.  We adopt a detection limit of $3\sigma^\mathrm{rms}_\nu$ to peak intensities at the nuclei.  In detected cases, signal-to-noise ratios, $S/N$, are 4.0--27 and 7.5--30 in the NMA and JCMT measurements, respectively.  Although a spurious emission of $\sim 3$--$4 \sigma^\mathrm{rms}_\nu$ sometimes appears in field of view of the NMA, we believe our detections because all of detected emission components are found as a single point source at just the phase-tracking centre to an accuracy of less than HPBW$/(S/N)$.  We set the phase-tracking centres where compact components had been found in the past VLBI-astrometric observations.  Because of $S/N>7.5$ there is also no doubt in the detections with the JCMT.  In undetected cases, the upper limit of flux densities takes $3 \sigma^\mathrm{rms}_\nu$.

In addition to the past observations in literatures, we finally obtain centimetre-submillimetre continuum radio spectra for the LLAGNs of our sample (Fig.~\ref{figure1}, and their references are listed in Table~\ref{table2}).

\begin{figure*}
\begin{center}
\begin{minipage}{160mm}
\includegraphics[width=\linewidth]{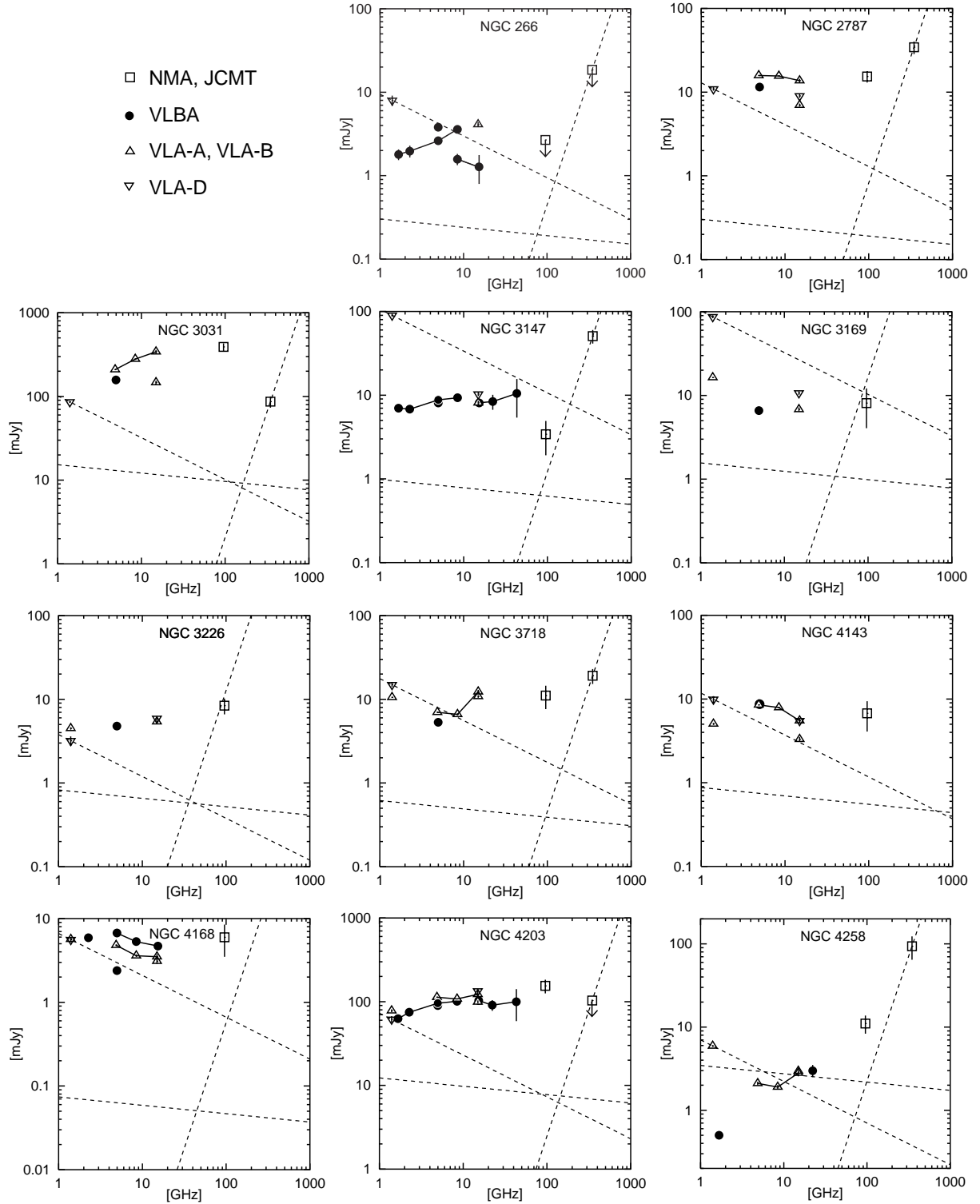}

\caption{Radio spectra of the VLBI-detected LLAGN sample.  The symbols connected with solid lines represent quasi-simultaneous observations.  For undetected data, $3\sigma^\mathrm{rms}_\nu$ upper limits are illustrated with downward arrows, where $\sigma^\mathrm{rms}_\nu$ is r.m.s.~of image noise.  Only for JCMT data at 347~GHz, intensities at nuclei, rather than total flux densities, are plotted.  Dashed lines represent putative spectra only for visualizing the upper limits of potential contributions to measured 96-GHz flux densities from diffuse nonthermal~($\propto \nu^{-0.5}$), free-free~($\propto \nu^{-0.1}$), and dust ($\propto \nu^{+3}$, where $\nu$ is frequency) components (see Section~\ref{section:diffusefraction} and Table~\ref{table3}).  Only for NGC 4143, no putative dust spectrum is not described because no data for estimation of dust contribution was available.  Free-free emissions in NGC~4374, NGC~4472, NGC~4486, and NGC~4552 are too weak to appear in the ranges of the plots.
}
\label{figure1}
\end{minipage}
\end{center}
\end{figure*}

\begin{figure*}
\begin{center}
\begin{minipage}{160mm}
\includegraphics[width=\linewidth]{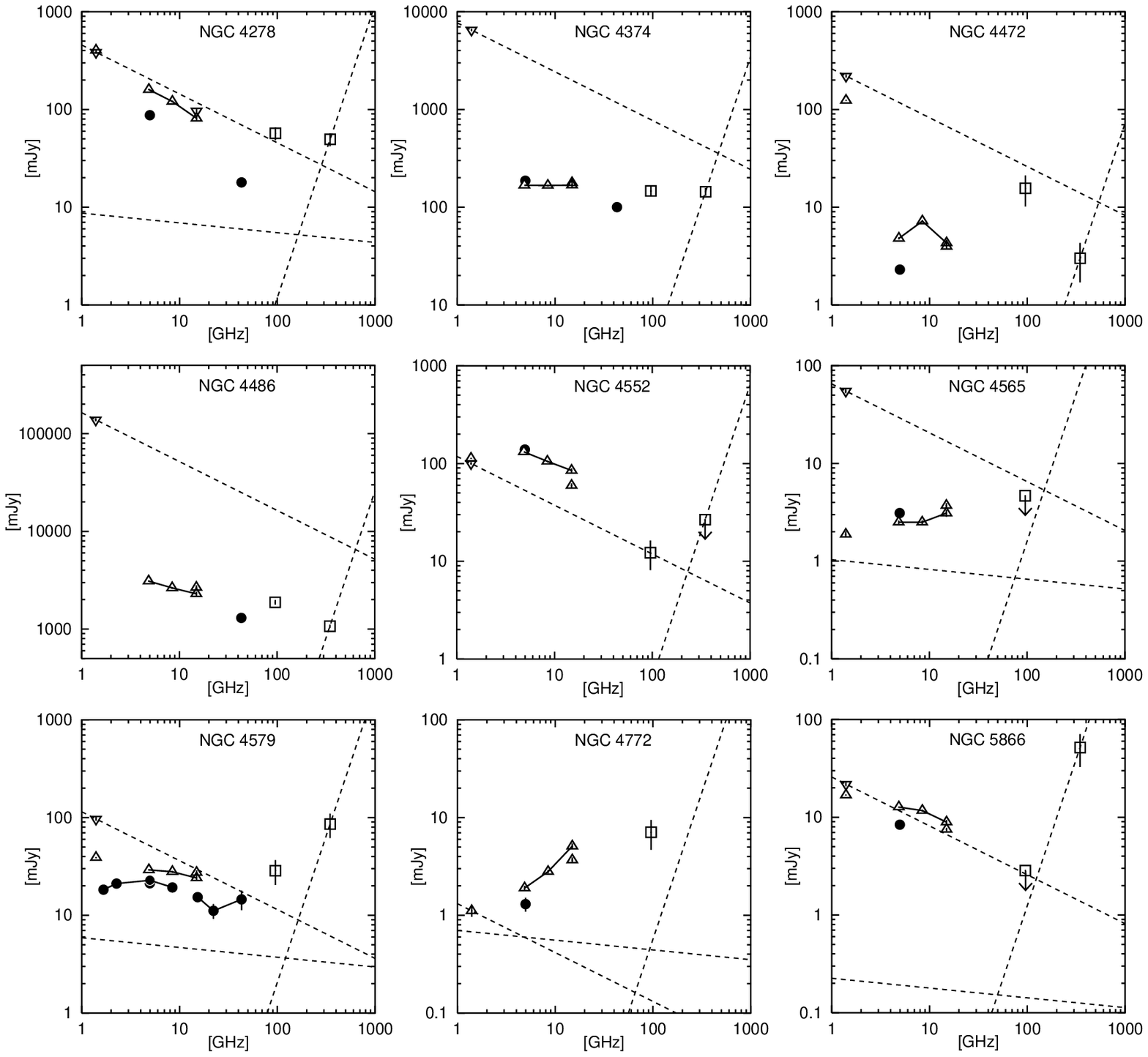}
\contcaption{}
\end{minipage}
\end{center}
\end{figure*}

\begin{table}
\caption{Reference list for Fig.~1.}

\defcitealias{Condon_etal.1998}{a}
\defcitealias{Nagar_etal.2000}{b}
\defcitealias{Nagar_etal.2001}{c}
\defcitealias{Nagar_etal.2002}{d}
\defcitealias{Becker_etal.1995}{e}
\defcitealias{Falcke_etal.2000}{f}
\defcitealias{Bartel_Bietenholz2000}{g}
\defcitealias{Ulvestad_Ho2001a}{h}
\defcitealias{Doi_etal.2005}{i}
\defcitealias{Anderson_etal.2004}{j}
\defcitealias{Herrnstein_etal.1997}{k}
\defcitealias{Cecil_etal.2000}{l}
\defcitealias{Ly_etal.2004}{m}
\defcitealias{Fomalont_etal.2000}{n}
\defcitealias{Sakamoto_etal.2001}{o}
\defcitealias{Leeuw_etal.2000}{q}
\defcitealias{diMatteo_etal.1999}{r}
\defcitealias{Despringre_etal.1996}{s}
\defcitealias{Knapp_Patten1991}{t}

\begin{tabular}{lllll} 
\hline
\multicolumn{1}{c}{Name} & \multicolumn{1}{c}{VLA-D} & \multicolumn{1}{c}{VLA-A/} & \multicolumn{1}{c}{VLBI} & \multicolumn{1}{c}{NMA/} \\
\multicolumn{1}{c}{} & \multicolumn{1}{c}{} & \multicolumn{1}{c}{VLA-B} & \multicolumn{1}{c}{} & \multicolumn{1}{c}{JCMT} \\
\multicolumn{1}{c}{(1)} & \multicolumn{1}{c}{(2)} & \multicolumn{1}{c}{(3)} & \multicolumn{1}{c}{(4)} & \multicolumn{1}{c}{(5)} \\\hline
NGC 266 & \citetalias{Condon_etal.1998} & \citetalias{Nagar_etal.2000} & \citetalias{Falcke_etal.2000}i & i \\
NGC 2787 & \citetalias{Condon_etal.1998}\citetalias{Nagar_etal.2000} & \citetalias{Nagar_etal.2000}\citetalias{Nagar_etal.2001} & \citetalias{Falcke_etal.2000} & x \\
NGC 3031 & \citetalias{Condon_etal.1998} & \citetalias{Nagar_etal.2002}\citetalias{Nagar_etal.2001} & \citetalias{Bartel_Bietenholz2000} & \citetalias{Sakamoto_etal.2001}$^{*1}$ \\
NGC 3147 & \citetalias{Condon_etal.1998}\citetalias{Nagar_etal.2000} & \citetalias{Nagar_etal.2000} & \citetalias{Ulvestad_Ho2001a}\citetalias{Anderson_etal.2004} & x \\
NGC 3169 & \citetalias{Condon_etal.1998}\citetalias{Nagar_etal.2000} & \citetalias{Becker_etal.1995}\citetalias{Nagar_etal.2000} & \citetalias{Falcke_etal.2000} & x \\
NGC 3226 & \citetalias{Condon_etal.1998}\citetalias{Nagar_etal.2000} & \citetalias{Becker_etal.1995}\citetalias{Nagar_etal.2000} & \citetalias{Falcke_etal.2000} & x \\
NGC 3718 & \citetalias{Condon_etal.1998} & \citetalias{Becker_etal.1995}\citetalias{Nagar_etal.2002}\citetalias{Nagar_etal.2001} & \citetalias{Nagar_etal.2002} & x \\
NGC 4143 & \citetalias{Condon_etal.1998}\citetalias{Nagar_etal.2000} & \citetalias{Becker_etal.1995}\citetalias{Nagar_etal.2000}\citetalias{Nagar_etal.2001} & \citetalias{Nagar_etal.2002} & x \\
NGC 4168 & \citetalias{Condon_etal.1998} & \citetalias{Becker_etal.1995}\citetalias{Nagar_etal.2002}\citetalias{Nagar_etal.2001} & \citetalias{Nagar_etal.2002}\citetalias{Anderson_etal.2004} & x \\
NGC 4203 & \citetalias{Condon_etal.1998}\citetalias{Nagar_etal.2000} & \citetalias{Becker_etal.1995}\citetalias{Nagar_etal.2000}\citetalias{Nagar_etal.2001} & \citetalias{Falcke_etal.2000}\citetalias{Ulvestad_Ho2001a}\citetalias{Anderson_etal.2004} & x \\ 
NGC 4258 &  & \citetalias{Becker_etal.1995}\citetalias{Nagar_etal.2002}\citetalias{Nagar_etal.2001} & \citetalias{Herrnstein_etal.1997}\citetalias{Cecil_etal.2000} & p$^{*1}$ \\
NGC 4278 & \citetalias{Condon_etal.1998}\citetalias{Nagar_etal.2000} & \citetalias{Becker_etal.1995}\citetalias{Nagar_etal.2000}\citetalias{Nagar_etal.2001} & \citetalias{Falcke_etal.2000}\citetalias{Ly_etal.2004} & x \\
NGC 4374 & \citetalias{Condon_etal.1998} & \citetalias{Nagar_etal.2002}\citetalias{Nagar_etal.2001} & \citetalias{Nagar_etal.2002}\citetalias{Ly_etal.2004} & x\citetalias{Leeuw_etal.2000}$^{*1}$ \\
NGC 4472 & \citetalias{Condon_etal.1998} & \citetalias{Becker_etal.1995}\citetalias{Nagar_etal.2002}\citetalias{Nagar_etal.2001} & \citetalias{Nagar_etal.2002} & x\citetalias{diMatteo_etal.1999} \\
NGC 4486 & \citetalias{Condon_etal.1998} & \citetalias{Nagar_etal.2002}\citetalias{Nagar_etal.2001} & \citetalias{Fomalont_etal.2000}\citetalias{Ly_etal.2004} & \citetalias{Despringre_etal.1996}$^{*2}$\citetalias{Knapp_Patten1991} \\
NGC 4552 & \citetalias{Condon_etal.1998} & \citetalias{Becker_etal.1995}\citetalias{Nagar_etal.2002}\citetalias{Nagar_etal.2001} & \citetalias{Nagar_etal.2002} & x \\
NGC 4565 & \citetalias{Condon_etal.1998} & \citetalias{Becker_etal.1995}\citetalias{Nagar_etal.2000}\citetalias{Nagar_etal.2001} & \citetalias{Falcke_etal.2000} & x \\
NGC 4579 & \citetalias{Condon_etal.1998} & \citetalias{Becker_etal.1995}\citetalias{Nagar_etal.2000}\citetalias{Nagar_etal.2001} & \citetalias{Falcke_etal.2000}\citetalias{Anderson_etal.2004}\citetalias{Ulvestad_Ho2001a} & x \\
NGC 4772 &  & \citetalias{Becker_etal.1995}\citetalias{Nagar_etal.2002}\citetalias{Nagar_etal.2001} & \citetalias{Nagar_etal.2002} & x \\
NGC 5866 & \citetalias{Condon_etal.1998} & \citetalias{Becker_etal.1995}\citetalias{Nagar_etal.2000}\citetalias{Nagar_etal.2001} & \citetalias{Falcke_etal.2000} & x \\\hline
\end{tabular}
\label{table2}

\textbf{*1} --- mean flux density of multi-epoch data.\\
\textbf{*2} --- data obtained with the IRAM Plateau de Bure interferometer at 89 GHz.

\smallskip
Note. --- Column are: \textbf{(1)} galaxy name; \textbf{(2)} reference of data obtained with the VLA D~configuration at 1.4 or 15~GHz; \textbf{(3)} reference of data obtained with the VLA A or B~configuration at 1.4-15~GHz; \textbf{(4)} reference of data obtained with VLBI; \textbf{(5)} reference of data obtained with the NMA and JMCT.

\smallskip
Note. --- Reference:
\textbf{\citetalias{Condon_etal.1998}}: \citealt{Condon_etal.1998};
\textbf{\citetalias{Nagar_etal.2000}}: \citealt{Nagar_etal.2000};
\textbf{\citetalias{Nagar_etal.2001}}: \citealt{Nagar_etal.2001};
\textbf{\citetalias{Nagar_etal.2002}}: \citealt{Nagar_etal.2002};
\textbf{\citetalias{Becker_etal.1995}}: \citealt{Becker_etal.1995};
\textbf{\citetalias{Falcke_etal.2000}}: \citealt{Falcke_etal.2000};
\textbf{\citetalias{Bartel_Bietenholz2000}}: \citealt{Bartel_Bietenholz2000};
\textbf{\citetalias{Ulvestad_Ho2001a}}: \citealt{Ulvestad_Ho2001a};
\textbf{\citetalias{Doi_etal.2005}}: \citealt{Doi_etal.2005};
\textbf{\citetalias{Anderson_etal.2004}}: \citealt{Anderson_etal.2004};
\textbf{\citetalias{Herrnstein_etal.1997}}: \citealt{Herrnstein_etal.1997};
\textbf{\citetalias{Cecil_etal.2000}}: \citealt{Cecil_etal.2000};
\textbf{\citetalias{Ly_etal.2004}}: \citealt{Ly_etal.2004};
\textbf{\citetalias{Fomalont_etal.2000}}: \citealt{Fomalont_etal.2000};
\textbf{\citetalias{Sakamoto_etal.2001}}: \citealt{Sakamoto_etal.2001};
\textbf{p}: Doi et al. in prep.;
\textbf{\citetalias{Leeuw_etal.2000}}: \citealt{Leeuw_etal.2000};
\textbf{\citetalias{diMatteo_etal.1999}}: \citealt{diMatteo_etal.1999};
\textbf{\citetalias{Despringre_etal.1996}}: \citealt{Despringre_etal.1996};
\textbf{\citetalias{Knapp_Patten1991}}: \citealt{Knapp_Patten1991};
\textbf{x}: the present paper
\end{table}

\subsection{Radio morphology}
In the NMA observations, all of the detected sources are found as a single radio component.  No significant extended emission was found in the images with the NMA beam of $\sim7$~arcsec.  This indicates that there is no detectable emission of extended component in outer regions of $\ga500$~pc from the nuclei.  However, some unidentified components other than the compact core could contribute into the NMA beam.  This effect will be evaluated in Section~\ref{section:diffusefraction}.

\begin{figure*}
\begin{center}
\begin{minipage}{170mm}
\bigskip
\includegraphics[width=\linewidth]{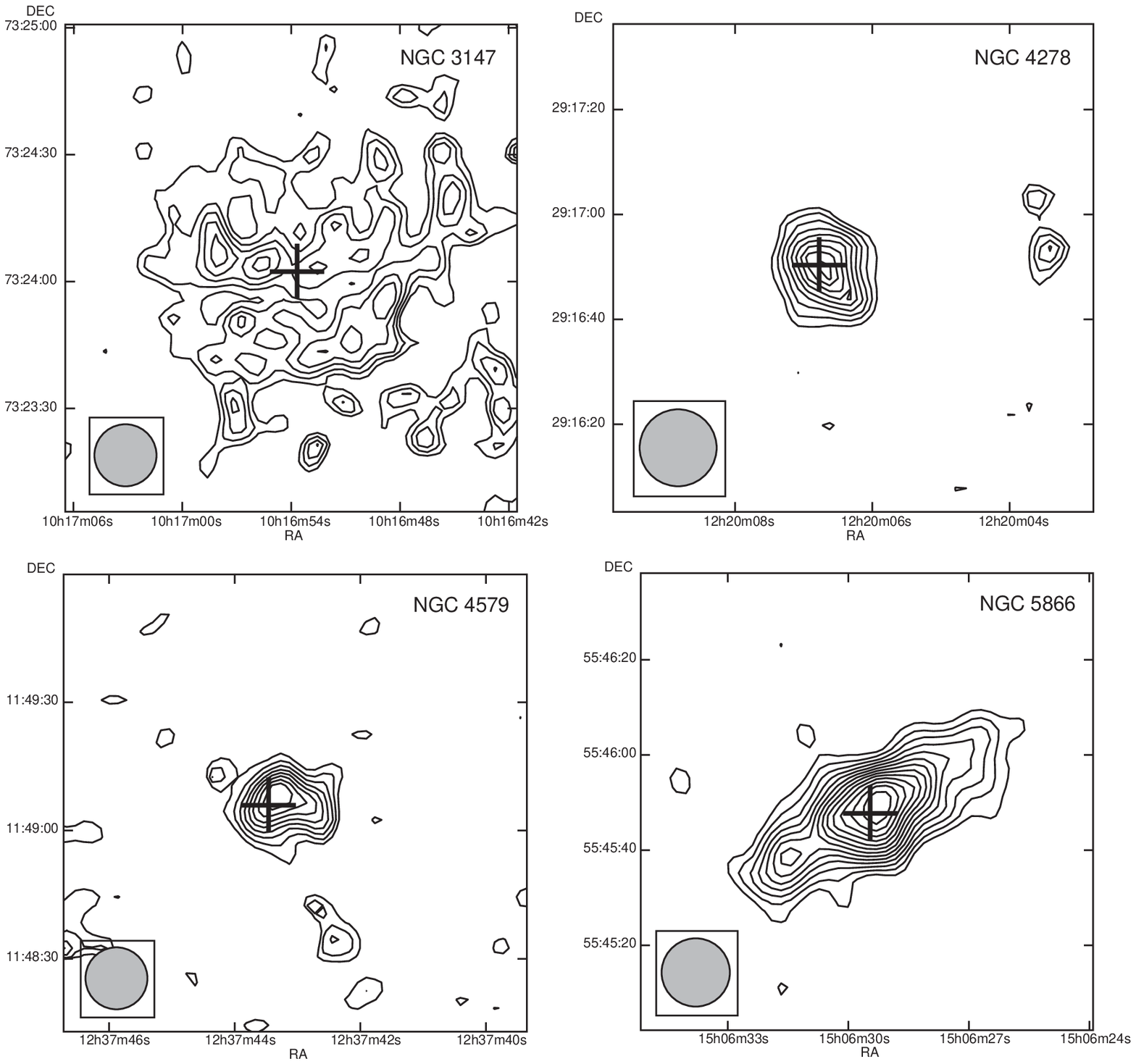}

\caption{JCMT images at 347~GHz for all the VLBI-detected LLAGNs that had been observed in SCUBA jiggle-map mode, except for NGC~266 \citep{Doi_etal.2005}, NGC~4258 (Doi~et~al. in prep.), and NGC~4374 \citep{Leeuw_etal.2000}.  Contour levels: NGC~3147, $\sigma^\mathrm{rms}_\mathrm{347GHz}\times$5,6,7, and 8; NGC~4278, $\sigma^\mathrm{rms}_\mathrm{347GHz}\times$3,4,5,\ldots and 11; NGC~4579, $\sigma^\mathrm{rms}_\mathrm{347GHz}\times$4,5,6,\ldots and 11; NGC~5866, $\sigma^\mathrm{rms}_\mathrm{347GHz}\times$3,4,5,\ldots and~15, where $\sigma^\mathrm{rms}_\mathrm{347GHz}$ is r.m.s. of image noise.  The HPBWs of the JCMT, 15.1~arcsec, are illustrated at the lower left-hand corner of the images.  Astrometric core positions are described in cross symbols.  NGC~3147, NGC~4579, and NGC~5866 are spatially resolved.}
\label{figure2}
\end{minipage}
\end{center}
\end{figure*}

On the other hand, in the JCMT measurements at 347~GHz, we found significant elongated features in NGC~4258 (Doi~et~al. in preparation), NGC~4579, and NGC~5866, and a very diffuse structure in NGC~3147 (Fig.~\ref{figure2}).  These features are consistent with the morphology of CO emissions that had been imaged with millimetric interferometers (NGC~4258, \citealt{Helfer_etal.2003}; NGC~4579, \citealt{Sofue_etal.2003}), with dust lane along the major axis of the host galaxy (NGC~5866, e.g., \citealt{Malhotra_etal.2000}), or with morphology of a low-frequency radio image (NGC 3147, \citealt{Hummel_etal.1985}).  It is, therefore, plausible that most of the 347-GHz emission from these sources originates in stellar processes of star-forming regions, probably dust emission, associated with their host galaxies.

\subsection{Spectral indices of a centimetre-to-millimetre band}\label{subsection:mmspectra}
Two-point spectral indices between 15~GHz (2~cm) and 96~GHz (3~mm), which had never been investigated, are derived in this section.  
Unmatched beam sizes at different frequencies leads to incorrect spectral-index measurements, if significant diffuse components are present (a `resolution effect').  Extended jets are one of the candidates.  Because of their power-law spectra this sort of contamination would be small at millimetre, but possible at centimetre wavelengths if we use a large beam.  
Hence, as the lower-frequency data to derive the two-point spectral indices, we use data from \citet{Nagar_etal.2000,Nagar_etal.2002} of the Very Large Array~(VLA) A-configuration at 15~GHz with a resolution of $\sim0.15$ arcsec, corresponding to $\sim14$~pc at 18.9~Mpc, which is a mean distance of our sample.  These are unique and uniform data set that can provide centimetre flux densities at the highest frequency and in the highest spatial resolution for all our LLAGN sample.  The other contaminations such as dust emission and free--free emission from \mbox{H\,{\sc ii}} regions could contribute especially to the NMA beam.  Very careful estimations for these diffuse contributions will be described in Section~\ref{section:diffusefraction}.

The two-point radio spectra are listed in column~(6) and (7) in Table~\ref{table3}.  These results are from unoperated, measured flux densities at 15 and 96~GHz, for which the diffuse contaminations are not removed.  We adopt a threshold of $\alpha=0$ between inverted and steep spectra.  The threshold definitely distinguishes 13 out of 20~sources beyond the margins of error of their spectral indices.  The other 6 out of remaining 7~sources were marginally distinguished; the threshold lies within the margins of error of their spectral indices.  We classified them into four spectral categories: definitely inverted `I', marginally inverted `i', marginally steep `s', and definitely steep `S' spectra.  Consequently, we found inverted spectra in many LLAGNs.  However, a substantial fraction of the LLAGNs showed steep spectra.

\begin{table*}
\defcitealias{Tully1988}{a}
\defcitealias{Tonry_etal.2001}{b}
\defcitealias{Freedman_etal.1994}{c}
\defcitealias{Herrnstein_etal.1999}{d}
\defcitealias{Solanes_etal.2002}{e}
\defcitealias{Nagar_etal.2000}{f}
\defcitealias{Nagar_etal.2002}{g}
\begin{center}
\begin{minipage}{175mm}
\caption{Source parameters for the VLBI-detected LLAGN sample.}

\begin{tabular}{llrclrccrrrrr} 
\hline
\multicolumn{1}{c}{Name} & \multicolumn{1}{c}{Class} & \multicolumn{1}{c}{Dist.} & Ref. & \multicolumn{1}{c}{Morph.} & \multicolumn{1}{c}{$\alpha_\mathrm{96GHz}^\mathrm{15GHz}$} & Sp. & Ref. & \multicolumn{1}{c}{$S^\mathrm{dust}_\mathrm{max}$} & \multicolumn{1}{c}{$S^\mathrm{ff}_\mathrm{max}$} & \multicolumn{1}{c}{$S^\mathrm{nonth}_\mathrm{max}$} & \multicolumn{1}{c}{$f_\mathrm{max}^\mathrm{nonth}$} & \multicolumn{1}{c}{$f_\mathrm{max}^\mathrm{total}$} \\
\multicolumn{1}{c}{} & \multicolumn{1}{c}{} & \multicolumn{1}{c}{(Mpc)} &  & \multicolumn{1}{c}{} & \multicolumn{1}{c}{} &  & $S_\mathrm{15GHz}$ & \multicolumn{1}{c}{(mJy)} & \multicolumn{1}{c}{(mJy)} & \multicolumn{1}{c}{(mJy)} & \multicolumn{1}{c}{} & \multicolumn{1}{c}{} \\
\multicolumn{1}{c}{(1)} & \multicolumn{1}{c}{(2)} & \multicolumn{1}{c}{(3)} & (4) & \multicolumn{1}{c}{(5)} & \multicolumn{1}{c}{(6)} & (7) & (8) & \multicolumn{1}{c}{(9)} & \multicolumn{1}{c}{(10)} & \multicolumn{1}{c}{(11)} & \multicolumn{1}{c}{(12)} & \multicolumn{1}{c}{(13)} \\\hline
NGC 266 & L1.9 & 62.4  & \citetalias{Tully1988} & SB(rs)ab & $<-$0.23 & S & \citetalias{Nagar_etal.2000} & 0.4  & 0.2  & 1.0  & \ldots & \ldots \\
NGC 2787 & L1.9 & 7.5  & \citetalias{Tonry_etal.2001} & SB(r)0+ & 0.42$\pm$0.14 & I & \citetalias{Nagar_etal.2000} & 0.7  & 0.7  & 1.3  & 0.09  & 0.18  \\
NGC 3031 & S1.5 & 3.6  & \citetalias{Freedman_etal.1994} & SA(s)ab & 0.46$\pm$0.09 & I & \citetalias{Nagar_etal.2002} & 1.8  & 9.7  & 10.4  & 0.03  & 0.06  \\
NGC 3147 & S2 & 40.9  & \citetalias{Tonry_etal.2001} & SA(rs)bc & $-$0.47$\pm$0.50 & s & \citetalias{Nagar_etal.2000} & 1.1  & 0.6  & 10.9  & 3.18  & 3.68  \\
NGC 3169 & L2 & 19.7  & \citetalias{Tonry_etal.2001} & SA(s)a pec & 0.09$\pm$0.57 & i & \citetalias{Nagar_etal.2000} & 14.6$^{*1}$ & 1.0  & 10.5  & 1.30  & 3.22  \\
NGC 3226 & L1.9 & 23.6  & \citetalias{Tonry_etal.2001} & E2: pec & 0.24$\pm$0.21 & I & \citetalias{Nagar_etal.2000} & 11.1$^{*1}$ & 0.5  & 0.4  & 0.05  & 1.43  \\
NGC 3718 & L1.9 & 17.0  & \citetalias{Tully1988} & SB(s)a pec & 0.01$\pm$0.34 & i & \citetalias{Nagar_etal.2002} & 0.4  & 0.4  & 1.8  & 0.16  & 0.23  \\
NGC 4143 & L1.9 & 15.9  & \citetalias{Tonry_etal.2001} & SAB(s)0 & 0.39$\pm$0.41 & i & \citetalias{Nagar_etal.2000} & \ldots & 0.6  & 1.2  & 0.18  & \ldots \\
NGC 4168 & S1.9 & 30.9  & \citetalias{Tonry_etal.2001} & E2 & 0.35$\pm$0.43 & i & \citetalias{Nagar_etal.2002} & 0.5$^{*1}$ & 0.0  & 0.7  & 0.11  & 0.21  \\
NGC 4203 & L1.9 & 15.1  & \citetalias{Tonry_etal.2001} & SAB0-: & 0.26$\pm$0.19 & I & \citetalias{Nagar_etal.2000} & 0.2  & 0.8  & 0.7  & 0.05  & 0.11  \\
NGC 4258 & S1.9 & 7.3  & \citetalias{Herrnstein_etal.1999} & SAB(s)bc & 0.70$\pm$0.21 & I & \citetalias{Nagar_etal.2002} & 2.0  & 2.2  & 0.7  & 0.07  & 0.44  \\
NGC 4278 & L1.9 & 16.1  & \citetalias{Tonry_etal.2001} & E1+ & $-$0.25$\pm$0.15 & S & \citetalias{Nagar_etal.2000} & 1.0  & 5.5  & 46.6  & 0.82  & 0.93  \\
NGC 4374 & L2 & 18.4  & \citetalias{Tonry_etal.2001} & E1 & $-$0.12$\pm$0.11 & S & \citetalias{Nagar_etal.2002} & 3.0  & 1.0  & 785$^{*3}$ & 5.36  & 5.38  \\
NGC 4472 & S2 & 16.3  & \citetalias{Tonry_etal.2001} & E2 & 0.72$\pm$0.36 & I & \citetalias{Nagar_etal.2002} & 0.1  & 0.0  & 26.6  & 1.70  & 1.71  \\
NGC 4486 & L2 & 16.1  & \citetalias{Tonry_etal.2001} & E0 + pec & $-$0.22$\pm$0.03 & S & \citetalias{Nagar_etal.2002} & 22.4  & 3.4  & 16748  & 8.91  & 8.92  \\
NGC 4552 & T2 & 15.3  & \citetalias{Tonry_etal.2001} & E & $-$0.85$\pm$0.38 & S & \citetalias{Nagar_etal.2002} & 0.6  & 0.4  & 12.1  & 0.99  & 1.07  \\
NGC 4565 & S1.9 & 17.5  & \citetalias{Tonry_etal.2001} & SA(s)b?spin & $<$0.13 & \ldots & \citetalias{Nagar_etal.2000} & 1.4$^{*2}$ & 0.7  & 6.7  & \ldots & \ldots \\
NGC 4579 & S/L1.9 & 19.1  & \citetalias{Solanes_etal.2002} & SAB(rs)b & 0.01$\pm$0.32 & i & \citetalias{Nagar_etal.2000} & 1.8  & 3.7  & 11.7  & 0.41  & 0.60  \\
NGC 4772 & L1.9 & 16.3  & \citetalias{Tully1988} & SA(s)a & 0.39$\pm$0.34 & I & \citetalias{Nagar_etal.2002} & 0.5$^{*1}$ & 0.4  & 0.1  & 0.02  & 0.15  \\
NGC 5866 & T2 & 15.3  & \citetalias{Tonry_etal.2001} & SA0 + spin & $<-$0.52 & S & \citetalias{Nagar_etal.2000} & 1.1  & 0.1  & 2.6  & \ldots & \ldots \\\hline
\end{tabular}
\label{table3}

\textbf{*1} --- derived from {\it IRAS} data, see text in detail.\\
\textbf{*2} --- derived from 230-GHz data, see text in detail.\\
\textbf{*3} --- derived from data obtained with a single-dish telescope, see text in detail.

\smallskip
Note. --- Column are: \textbf{(1)} galaxy name; \textbf{(2)} nuclear activity type as determined in Palomar optical spectroscopic survey \citep{Ho_etal.1997a}, `L' represents LINER, `S' represents Seyfert, `T' represents objects with transitional L $+$ H\,{\sc ii} region-type spectrum.  `2' implies no broad H$\alpha$ is detected, `1.9' implies broad H$\alpha$ is present but not broad H$\beta$, and `1.5' implies that broad H$\alpha$ and broad H$\beta$ are detected; \textbf{(3)} distance to galaxy; \textbf{(4)} reference of the distance, as listed below; \textbf{(5)} Morphology type of host galaxy, as listed in \citet{Ho_etal.1997a}; \textbf{(6)} two-point spectral index between 15~GHz and 95~GHz, derived from data with the VLA A~configuration and NMA, respectively; \textbf{(7)} spectral category, classified by a threshold of $\alpha^\mathrm{15GHz}_\mathrm{96GHz}=0$.  `S' and `s' represent steep spectra, $\alpha^\mathrm{15GHz}_\mathrm{96GHz}<0$, determined definitely and marginally, respectively.  `I' and `i' represent inverted spectra, $\alpha^\mathrm{15GHz}_\mathrm{96GHz} \gid 0$, determined definitely and marginally, respectively; \textbf{(8)} reference for the VLA data at 15~GHz, as listed below; \textbf{(9)} upper limit of the potential contribution at 96~GHz from dust emission.  This was estimated mainly from peak intensity at 347~GHz and by extrapolating a dust spectrum, see text in detail; \textbf{(10)} upper limit of the potential contribution at 96~GHz from free--free emission.  This was estimated from H$\beta$ flux measured by \citet{Ho_etal.1997c} and \citet{Ho_etal.1997a}, see text in detail; \textbf{(11)} upper limit of the potential contribution at 96~GHz from diffuse nonthermal emission.  This was estimated from data at 1.4~GHz, see text in detail; \textbf{(12)} upper limit of diffuse fraction only of the nonthermal component in the 96~GHz flux density; \textbf{(13)} upper limit of total diffuse fraction in the 96~GHz flux density.

\smallskip
Note. --- Reference:  
\textbf{\citetalias{Tully1988}}: \citealt{Tully1988};
\textbf{\citetalias{Tonry_etal.2001}}: \citealt{Tonry_etal.2001};
\textbf{\citetalias{Freedman_etal.1994}}: \citealt{Freedman_etal.1994};
\textbf{\citetalias{Herrnstein_etal.1999}}: \citealt{Herrnstein_etal.1999};
\textbf{\citetalias{Solanes_etal.2002}}: \citealt{Solanes_etal.2002};
\textbf{\citetalias{Nagar_etal.2000}}: \citealt{Nagar_etal.2000};
\textbf{\citetalias{Nagar_etal.2002}}: \citealt{Nagar_etal.2002}
\end{minipage}
\end{center}
\end{table*}

\subsection{Submillimetre emission}
Submillimetre spectra of compact cores are a very interesting issue for observational examination of the ADAF model.  However, the JCMT flux densities could be dominated by dust emission from host galaxies, since beam widths of the JCMT were too large.  If the core was brighter than the dust, it may be detectable from its flux variation even with the large beam.  NGC~4374, NGC~4472, NGC~4579, and NGC~5866 had been observed at multiple epochs with the JCMT.  A significant variability was seen in NGC~4579: $85.7\pm23.8$~mJy~beam$^{-1}$ on 2002 January~16 by jiggle-map mode and $25.8\pm4.9$~mJy~beam$^{-1}$ on 2001 March~15 by photometry mode.  The significant fraction of observed submillimetre fluxes may be direct emission of the core.  In the other sources, no significant variation was found.  At least, the results from the JCMT measurements should be useful for estimation of the upper limits of dust contribution (Section~\ref{subsubsection:DustContamination})

\subsection{Diffuse fractions in 96-GHz flux densities}\label{section:diffusefraction}
The beam mismatch between the measurements of 15 and 96~GHz might pollute the two-point spectral indices.  Because the NMA beam at 96~GHz, $\sim7$ arcsec, was greatly larger than the VLA one at 15~GHz, $\sim0.15$ arcsec, the derived spectra would be biased toward being inverted if some diffuse components exist in addition to the core.  Possible contaminations are (1) dust emission, (2) free--free emission from plasma heated/ionized by UV radiation from OB stars in nuclear star-forming regions or an accretion disc of LLAGN, and (3) extended non-thermal emission from jets or supernova remnants.  In the next three subsections, we estimate the contaminations from the three diffuse components to the NMA measurements.

\subsubsection{Contamination from dust emission}\label{subsubsection:DustContamination}
The dust emission at a given frequency can be estimated by extrapolating an appropriate spectrum that was determined by measurements with the same aperture at higher frequencies where the dust contribution predominates any other components.  The graybody spectrum of dust emission is \mbox{$S_\nu^\mathrm{dust}\propto\nu^{2+\beta}$} in the Rayleigh-Jeans regime at less than about 1~THz, where $\beta$ is the dust emissivity, which is believed to be between 1 and 2 \citep{Hildebrand1983}.  From the JCMT results at 347~GHz, we can derive useful upper limits of the dust contribution to the observed 96-GHz flux densities by following three overestimates.

(i) The spectral indices of the dust spectra differ from galaxy to galaxy.  The $\beta$ value has been determined to be $1.3\pm0.2$ by JCMT observations for nearby galaxies \citep{Dunne_etal.2000}.  To keep it safe, we adopt $\beta=1$, which is the most influential case in the spectral extrapolation to lower frequencies.  This must provide an overestimate for dust contribution at 96~GHz.  (ii) Beam size of the JCMT, 15.1~arcsec, is about two times larger than that of the NMA, so it must lead to also an overestimate as photometry for the NMA aperture.  (iii) Contributions from the other components such as a core, jets, \mbox{H\,{\sc ii}} regions, and molecular clouds [CO($J=3-2$) lines at 345.796~GHz] etc., could contribute the JCMT measurements, so it must also make an overestimate compared to a measurement only for dust emission.   

Thus, from the three overestimates, we can derive an upper limit of the contamination from dust in the NMA measurement for each LLAGN.  For sources without the JCMT measurement, we used flux-density data from literatures: a single-dish observation at 230~GHz with a 20-arcsec beam for NGC~4565 \citep{Neininger_etal.1996}; the {\it IRAS} Faint Source Catalogue \citep{Moshir_etal.1990} at 60 and 100~$\mathrm{\mu m}$ with $\sim$1--2~arcmin beams and the relation between far-infrared~(FIR) and 850-$\mathrm{\mu m}$ total fluxes \citep{Dunne_etal.2000} for NGC~3169, NGC~3226, NGC~4168, and NGC~4772.  Unfortunately, no appropriate measurement for NGC~4143 at higher frequencies than 96~GHz was available.  The upper limits of the flux densities of the diffuse dust emission at 96~GHz, $S^\mathrm{dust}_\mathrm{max}$, are listed in column (9) of Table~\ref{table3}.  These values suggest that there must be little influence to the NMA measurements for all the LLAGNs, except for a few sources that was evaluated using the {\it IRAS} data.  The {\it IRAS} beams may be too large only for the nuclear region.  Therefore, all the NMA measurements are probably little affected from dust contamination.  While we adopted the extreme of dust spectra with $\beta=1$, there is nevertheless the slim possibilities that we may have detected at 96~GHz an unusual dust component with an abundance of very cold or very small grains \citep[cf.][]{Lisenfeld_etal.2002}.

\subsubsection{Contamination from free--free emission}\label{subsubsection:FreeFreeContamination}

Free--free emission from \mbox{H\,{\sc ii}} regions in nuclear star-formation or narrow-/broad-line regions around the AGN can be independently estimated from the extinction-corrected flux of H$\beta$ line emission by assuming an electron temperature and metallicity of the plasma \citep{Caplan&Deharveng1986,Condon1992}.  If the electron temperature is $\sim 10^4$ K and $N(\mathrm{He^+})/N(\mathrm{H^+})\sim0.08$, 
\begin{equation}
\left( \frac{S_\nu^\mathrm{ff}}{\mathrm{mJy}} \right) \sim 3.57 \times 10^{12} \left( \frac{F(\mathrm{H\beta})}{\mathrm{erg\ cm^{-2}\ s^{-1}}} \right) \left( \frac{\nu}{\mathrm{GHz}} \right)^{-0.1}, 
\end{equation}
where $S^\mathrm{ff}_\nu$ is the expected flux density of free--free emission at frequency $\nu$ and $F(\mathrm{H\beta})$ is $H\beta$ line flux.  The extinction values and extinction-corrected fluxes of broad and narrow H$\beta$ for our sample can be derived from the data of the Palomar optical spectroscopic survey \citep{Ho_etal.1997a,Ho_etal.1997c} using the extinction curve of \citet{Cardelli_etal.1989}.  We can derive useful upper limits of the free--free contribution to the NMA measurements by following two overestimates.

(i) Amount of hidden broad line components have been unknown in the optical survey.  In its sample, fractions of broad line flux in total H$\alpha$ flux are at most 0.98 \citep{Ho_etal.1997c}, which may be the least hidden and the most broad-line-dominant case.  We adopted this fraction, then the upper limit of the broad line fluxes can be derived using extinction-corrected narrow line fluxes.  It may be an overestimate of an intrinsic total line flux.  (ii) This optical survey was done with $2\times4$ arcsec$^{2}$ aperture, $\sim4.8$ times smaller than that of the NMA 7-arcsec beam.  Hence, we magnified the extinction-corrected narrow line H$\beta$ flux from a $2\times4$ arcsec$^2$ aperture to 4.8 times, then obtained an equivalent flux to the 7-arcsec region.  It can provide an overestimate because we assumed here that the surface density of narrow-line clouds is constant over the 7-arcsec region, although that tends to be rather highly central concentrated \citep{Pogge_etal.2000}.  We dealt here with only narrow-line components, which could be distributed over about a few hundred parsecs from the nucleus, while broad-line ones should concentrate within a sub-parsec region.  

From the two overestimates, we can derive the upper limit of the free--free emission at 96~GHz, $S^\mathrm{ff}_\mathrm{max}$, in the NMA measurement for each LLAGN.  The results, listed in column (10) of Table~\ref{table3}.  It suggests that there is little influence to the NMA measurements.

\subsubsection{Contamination from extended non-thermal emission} \label{subsubsection:nonthermal}

The extended non-thermal components of jets or supernova remnants must be optically thin, so show steep spectra with $\alpha \lid -0.5$.  A lower-frequency observation with a larger beam should be more sensitive to this emission.  With published data that had been obtained with larger beams at 1.4 GHz, we can deduce useful upper limits of contamination from the extended non-thermal components to the NMA measurements by following the two overestimates.

(i) We use data of the NRAO VLA Sky Survey (NVSS; \citealt{Condon_etal.1998}) at 1.4~GHz with $\sim45$-arcsec beam, which is significantly larger than that of the NMA.  For two sources, NGC~4258 and NGC~4772, uncovered in the NVSS, we used data of the Faint Images of the Radio Sky at Twenty-centimetre (FIRST) survey with $\sim5$-arcsec beam.  The beam of the FIRST survey was slightly smaller than that of the NMA, but emission with a size of $\sim7$~arcsec was expected to be detectable without being resolved out. Only NGC~4374 was uncovered in both the catalogues, therefore we used data of \citet{White&Becker1992} with $\sim700$-arcsec beam.  These 1.4~GHz measurements give us estimations of amount of extended nonthermal components in roughly the same or larger region than the NMA beam.  It makes an overestimate of a nonthermal flux into the NMA beam.  (ii) We adopt $\alpha=-0.5$ that would give the most influential case by extrapolating a power-law spectrum from 1.4 to 96~GHz.  It makes an overestimate of a nonthermal flux density at 96~GHz.

From the two overestimates, we can deduce an upper limit of the extended nonthermal contamination, $S^\mathrm{nonth}_\mathrm{max}$, in the NMA measurement for each source.  The results are listed in column (11) of Table~\ref{table3}.  For about half of our sample, very large contamination is possible.  Most of them show jet-like structures, which have been found by past observations (Sections~\ref{section:origin} and~\ref{section:individual}).  For the other half, little contamination is ensured.

\subsubsection{Total diffuse contaminations and spectral~indices}\label{section:spect-diffuse}

For our NMA measurements, we obtain the upper limits of possible contaminating flux densities at 96~GHz, $S^\mathrm{dust}_\mathrm{max}$, $S^\mathrm{ff}_\mathrm{max}$, and $S^\mathrm{nonth}_\mathrm{max}$, from dust, free--free, and nonthermal emission components, respectively.  The upper limit of total contamination fraction is now defined as $f^\mathrm{total}_\mathrm{max} \equiv [S^\mathrm{dust}_\mathrm{max} + S^\mathrm{ff}_\mathrm{max} + S^\mathrm{nonth}_\mathrm{max}] / S_\mathrm{96GHz}$.  The derived values are listed in column (13) of Table~\ref{table3}.  The value of $S^\mathrm{nonth}_\mathrm{max}$ accounts for the major part of $f^\mathrm{total}_\mathrm{max}$ for almost all the LLAGNs, except for a few sources whose dust contributions was estimated using the {\it IRAS} data with the much larger beam sizes.  This can be seen by comparing $f^\mathrm{nonth}_\mathrm{max}$ with $f^\mathrm{total}_\mathrm{max}$, where $f^\mathrm{nonth}_\mathrm{max} \equiv S^\mathrm{nonth}_\mathrm{max} / S_\mathrm{96GHz}$ listed in column (12) of Table~\ref{table3}.  It seems that both the diffuse dust and free-free emission components are practically negligible in arcsec-resolution observations at $\sim3$~mm for LLAGNs.  It is only necessary to concern that the diffuse nonthermal components could be responsible for artificially-made inverted spectra between 15 and 96~GHz due to unmatched resolutions; the LLAGNs with large $f^\mathrm{nonth}_\mathrm{max}$ would tend to be biased toward inverted spectra.

\begin{figure}
\includegraphics[width=\linewidth]{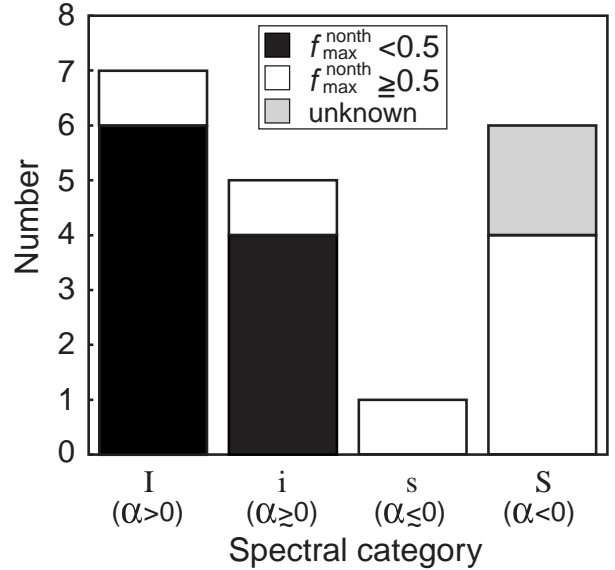}
\caption{Histogram of two-point spectral indices between 15 and 96~GHz, $\alpha^\mathrm{15GHz}_\mathrm{96GHz}$, for the VLBI-detected LLAGN sample.  Filled region represents sources with $f^\mathrm{nonth}_\mathrm{max} < 0.5$, where $f^\mathrm{nonth}_\mathrm{max} \equiv S^\mathrm{nonth}_\mathrm{max} / S_\mathrm{96GHz}$ (see Section~\ref{subsubsection:nonthermal} and \ref{section:spect-diffuse}, and Table~\ref{table3}).  It is warranted that diffuse contamination is practically negligible for these sources.  Open region represents sources with $f^\mathrm{nonth}_\mathrm{up} \geq 0.5$; we cannot ruled out significant diffuse contamination for these sources.  Grey region represent sources whose diffuse fractions cannot be estimated because of undetection in their NMA measurements.  NGC~4565 is not included here because of its deficient determination of spectral category.  This histogram confirms that there are many LLAGNs showing inverted spectra that are not artificially made by resolution effect.}
\label{figure3}
\end{figure}

However, our results do indicate rather an inverse tendency.  Almost all the LLAGN showing inverted spectra have a practically negligible diffuse contamination with $f^\mathrm{nonth}_\mathrm{max}<0.5$, while almost all the LLAGNs showing steep spectra have possibly large contaminations with $f^\mathrm{nonth}_\mathrm{max}\geq 0.5$ (Fig.~\ref{figure3}).  A Wilcoxon test about the spectral indices between the LLAGNs with $f^\mathrm{nonth}_\mathrm{max} < 0.5$ and those with $f^\mathrm{nonth}_\mathrm{max} \geq 0.5$ resulted in a probability of 0.0246 for the null hypothesis that these two groups were drawn from the same parent population.  Note that three LLAGNs whose $f^\mathrm{nonth}_\mathrm{max}$ or $\alpha^\mathrm{15GHz}_\mathrm{96GHz}$ values cannot had been be determined were excluded in this statistical test.  

Thus, for almost all the LLAGNs showing inverted spectra, we cannot find any evidence that the observed two-point spectra between 15 and 96~GHz had been artificially made by resolution effect.  The observed spectra presumably reflect the intrinsic spectra of the cores that had been detected as unresolved sources with the VLA in a 0.15-arcsec resolution.  Also, possible flux variability between observational epochs at 15 and 96~GHz could not have artificially made such many inverted spectra unless the spectra were intrinsically inverted; the LLAGNs sometimes show rapid variability \citep [e.g.,][]{Ho_etal.1999,Sakamoto_etal.2001,Nagar_etal.2002}, but our result must be statistically supported by a large number of our sample.

\section{Disscussion}\label{section:discussion}

\subsection{Origins of radio spectra}\label{section:origin}
We found inverted high-frequency radio spectra in a number of LLAGNs with compact cores.  On the other hand, a significant fraction of our sample showed steep spectra (Section~\ref{subsection:mmspectra}).  We discuss the variety of the observed radio spectra.  An interpretation is that optically-thin nonthermal jets may be responsible for these steep spectra, while the inverted spectra may be caused by ADAF discs.  The presence of jets are supposed to be recognized by very high radio powers that cannot be generated in the accretion disc, or by its characteristic radio morphology.  In this section, we investigate the radio core power and radio morphology of our sample, in order to have clues to the origins of the observed radio spectra.

\begin{figure}
\includegraphics[width=\linewidth]{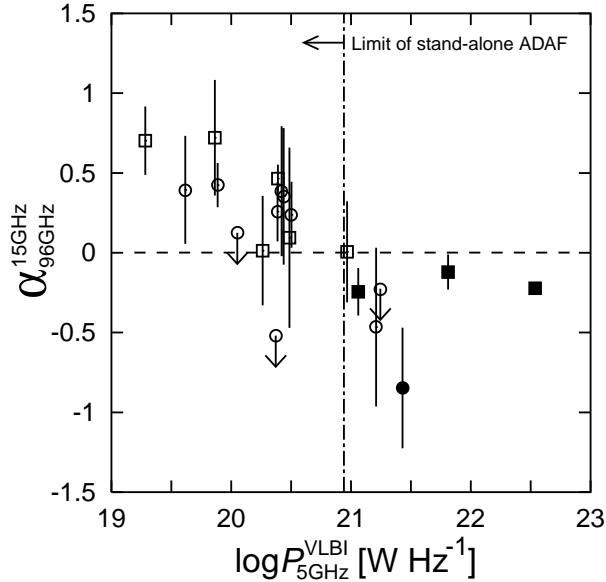}
\caption{Relation between 5-GHz radio power of core detected with VLBIs and spectral index between 15 and 96 GHz for the VLBI-detected LLAGN sample.  Squares represent sources with kpc-scale jets, while circles represent sources without kpc-scale jets.  Filled ones represent sources with pc-scale jets (Table~\ref{table4}).  The dot-dashed line represents the radio power limit that the ADAF on a black hole with $10^9$~M$_{\sun}$ can generate.}
\label{figure4}
\end{figure}

The relation between the two-point spectral index~($\alpha^\mathrm{15GHz}_\mathrm{96GHz}$) and the 5-GHz core power~($P^\mathrm{VLBI}_\mathrm{5GHz}$) that had been measured in $\sim1$~milli-arcsec resolutions of VLBIs (column [2] in Table~\ref{table4}) is shown in Fig.~\ref{figure4}.  The information about jet morphology (column [3] and [5] in Table~\ref{table4}) is also included in this plot.  From this plot, we can find the following three properties: (1)~a clear inverse correlation, (2)~there is no LLAGN showing inverted spectrum in the region of high core power, and (3)~the LLAGNs with pc-scale jets are found exclusively in the region of steep spectrum and higher core power.

An inverse correlation between the spectral index and VLBI core power can be clearly seen as Fig.~\ref{figure4}.  The LLAGNs with higher VLBI-core powers tend to show steeper spectra.  A generalized Kendall's $\tau$ test, using the ASURV statistical package \citep{Isobe_Feigelson1990,LaValley_etal.1992}, said that a probability is only 0.0023 for the null hypothesis that there is no correlation between them.  This suggests that the observed spectra are strongly related to the VLBI core powers.  The VLBI core powers are expected to include both the contributions from compact jets and the accretion disc.  However, because of the relatively low frequency, 5~GHz, the core power is expected to be biased toward the jet contribution.  On the other hand, because of relatively high frequencies, the spectral index may be biased toward the disc contribution.  The inverse correlation might appear as a consequence of the combination of such biases.  In other words, the inverse correlation suggests an intrinsic nature of LLAGNs: jets and disc coexist in the core \citep[a jet-ADAF model, e.g.,][]{Yuan_etal.2002}; the observed steep spectra are as the result of jet domination, and the observed inverted spectra are as the result of disc domination.

There is no LLAGNs showing inverted spectra in the region of high core power that cannot be generated by the ADAF.  In an simple ADAF model \citep{Mahadevan1997}, the maximum radio power at a given frequency is mainly dependent on black hole mass and accretion rate \citep[cf.~the same discussion in][]{Anderson_etal.2004}.  With a black hole mass of $10^9$~M$_{\sun}$, the ADAF cannot generate a 5-GHz radio power greater than $\sim8.9 \times 10^{20}$~W~Hz$^{-1}$ in any accretion rate (as a vertical line in Fig.\ref{figure4}).  It is consistent with the apparent upper limit of the VLBI core power of LLAGNs showing inverted spectra.  Hence, we can reasonably say that the ADAF may be responsible for the inverted spectra for the lower-core-power objects, while potential disc components may be dominated by jet components for the higher-core-power objects.

All the four LLAGNs with pc-scale jets are found exclusively in the region of steep spectrum and higher core power, as filled symbols in Fig.~\ref{figure4}.  Note that the 15-GHz flux densities, which were used for derivation of the two-point spectral indices, were from a region of $<$0.15 arcsec, typically corresponding to $\la14$~pc from the nuclei, where the pc-scale jets have been found.  It is, therefore, not surprising that the four LLAGNs were categorized as definitely-steep spectra `S' (Table~\ref{table4}).  Their high core powers are presumably generated by the base of these pc-scale jets, rather than discs.  On the other hand, six out of nine LLAGNs with kpc-scale jets reside in the range of inverted spectrum.  For four out of the six, we stress that their inverted spectra were not artificially made by resolution effect due to the kpc-scale jets because we have confirmed that the diffuse nonthermal flux densities into the NMA measurements were significantly small (Section~\ref{section:spect-diffuse}).  That is, the six LLAGNs have both compact cores showing inverted spectra and very faint kpc-scale jets.  Our NMA beam size was small enough to separate off their kpc-scale jets.  For the other two LLAGNs with kpc-scale jets, NGC~4472 and NGC~3169, we cannot rule out resolution effect.  The two LLAGNs are seen as the two open regions on the spectral category `I' and `i' in Fig.~\ref{figure3}, respectively.

Consequently, we propose an idea that the high-frequency radio spectra mainly depend on which is more dominant in the core, jets or disc.  This idea provides a reasonable explanation on the clear inverse correlation between the high-frequency radio spectrum and low-frequency core power.  The inverted spectra of the ADAF may be observable only when the jet components are relatively faint.  We may have detected direct radio emission from the accretion disc of the LLAGNs with faint jets.  The power ratios between disc and jet seem to differ from object to object.  It is very interesting for astrophysics on the issue of the accretion budget allocation on a black hole system.

However, we must note that our results do not rule out the possibility that the inverted spectra are made only by jets.  In a jet model proposed by \citet{Falcke_Biermann1999}, the submillimetre bump is produced by the acceleration zone of the jet, called nozzle, associated with the ADAF, while the low-frequency flat spectrum comes from the outer part of jet.  The jet model can explain the observed high-luminosities of low-frequency radio emission, which cannot be generated by a simple ADAF \citep{Ulvestad_Ho2001a,Yuan_etal.2002,Anderson_etal.2004}.  However, the outer jet structures must have been revealed in the VLBI images if the jet axes were not fairly close to our line of sight \citep{Anderson_etal.2004}.  To judge it, imaging analyses in micro-arcsec resolutions would be necessary.

\begin{table}
\caption{Monochromatic 5-GHz core power ranking and known jets in the VLBI-detected LLAGN sample.}
\defcitealias{Herrnstein_etal.1997}{a}
\defcitealias{Cecil_etal.2000}{b}
\defcitealias{Nagar_etal.2002}{c}
\defcitealias{Falcke_etal.2000}{d}
\defcitealias{Ulvestad_Ho2001a}{e}
\defcitealias{Anderson_etal.2004}{f}
\defcitealias{Bietenholz_etal.2000}{g}
\defcitealias{Bartel_Bietenholz2000}{h}
\defcitealias{Ly_etal.2004}{i}
\defcitealias{Giovannini_etal.2001}{j}
\defcitealias{Doi_etal.2005}{k}
\defcitealias{Fomalont_etal.2000}{l}
\defcitealias{Reid_etal.1989}{m}
\defcitealias{Junor_Biretta1995}{n}
\defcitealias{vanAlbada_vanderHulst1982}{o}
\defcitealias{Condon_Broderick1988}{p}
\defcitealias{Ho_Ulvestad2001}{q}
\defcitealias{Condon1987}{r}
\defcitealias{Kaufman_etal.1996}{s}
\defcitealias{Hummel_etal.1987}{t}
\defcitealias{Wrobel_Heeschen1984}{u}
\defcitealias{Laing_Bridle1987}{v}
\defcitealias{Owen_etal.1989}{w}

\begin{tabular}{lcclclc} 
\hline
\multicolumn{1}{c}{Name} & $\log{P^\mathrm{VLBI}_\mathrm{5GHz}}$ & Jet & \multicolumn{1}{c}{Ref.} & Jet & \multicolumn{1}{c}{Ref.} & Sp. \\
\multicolumn{1}{c}{} & (W Hz$^{-1}$) & VLBI & \multicolumn{1}{c}{} & VLA & \multicolumn{1}{c}{} &  \\
\multicolumn{1}{c}{(1)} & (2) & (3) & \multicolumn{1}{c}{(4)} & (5) & \multicolumn{1}{c}{(6)} & (7) \\\hline
NGC 4258 & 19.3  & S & \citetalias{Herrnstein_etal.1997}\citetalias{Cecil_etal.2000} & K & \citetalias{vanAlbada_vanderHulst1982} & I \\
NGC 4772 & 19.6  &  & \citetalias{Nagar_etal.2002} &  &  & I \\
NGC 4472 & 19.9  &  & \citetalias{Nagar_etal.2002} & K & \citetalias{Condon_Broderick1988}\citetalias{Ho_Ulvestad2001} & I \\
NGC 2787 & 19.9  &  & \citetalias{Falcke_etal.2000} &  &  & I \\
NGC 4565 & 20.1  &  & \citetalias{Falcke_etal.2000} &  &  & \ldots \\
NGC 3718 & 20.3  &  & \citetalias{Nagar_etal.2002} & K & \citetalias{Condon1987} & i \\
NGC 5866 & 20.4  &  & \citetalias{Falcke_etal.2000} &  &  & S \\
NGC 4203 & 20.4  &  & \citetalias{Falcke_etal.2000}\citetalias{Ulvestad_Ho2001a}\citetalias{Anderson_etal.2004} &  &  & I \\
NGC 3031 & 20.4  & S & \citetalias{Bietenholz_etal.2000}\citetalias{Bartel_Bietenholz2000} & K & \citetalias{Kaufman_etal.1996} & I \\
NGC 4143 & 20.4  &  & \citetalias{Nagar_etal.2002} &  &  & i \\
NGC 4168 & 20.4  &  & \citetalias{Nagar_etal.2002}\citetalias{Anderson_etal.2004} &  &  & i \\
NGC 3169 & 20.5  &  & \citetalias{Falcke_etal.2000} & K & \citetalias{Hummel_etal.1987} & i \\
NGC 3226 & 20.5  &  & \citetalias{Falcke_etal.2000} &  &  & I \\
NGC 4579 & 21.0  &  & \citetalias{Falcke_etal.2000}\citetalias{Anderson_etal.2004}\citetalias{Ulvestad_Ho2001a} & K & \citetalias{Ho_Ulvestad2001} & i \\
NGC 4278 & 21.1  & P & \citetalias{Falcke_etal.2000}\citetalias{Ly_etal.2004}\citetalias{Giovannini_etal.2001} & K & \citetalias{Wrobel_Heeschen1984} & S \\
NGC 3147 & 21.2  &  & \citetalias{Ulvestad_Ho2001a}\citetalias{Anderson_etal.2004} &  &  & s \\
NGC 266 & 21.2  &  & \citetalias{Falcke_etal.2000}k &  &  & S \\
NGC 4552 & 21.4  & P & \citetalias{Nagar_etal.2002} &  &  & S \\
NGC 4374 & 21.8  & P & \citetalias{Nagar_etal.2002}\citetalias{Ly_etal.2004} & K & \citetalias{Laing_Bridle1987} & S \\
NGC 4486 & 22.5  & P & \citetalias{Fomalont_etal.2000}\citetalias{Reid_etal.1989}\citetalias{Junor_Biretta1995} & K & \citetalias{Owen_etal.1989} & S \\\hline
\end{tabular}
\label{table4}

\smallskip
Note. --- Column are: \textbf{(1)} galaxy name; \textbf{(2)} core power detected in VLBI observations at 5 GHz, except for NGC~4258 at 22~GHz; \textbf{(3)} presence of jet feature in the VLBI image.  `S' represents sub-pc-scale jets.  `P' represents pc-scale jets; \textbf{(4)} reference for the VLBI observation, as listed below; \textbf{(5)} presence of jet feature in VLA image.  `K' represents kpc-scale jets; \textbf{(6)} reference for the VLA observation, as listed below; \textbf{(7)} spectral category, the same as Table~\ref{table3}.  The sources showing steep spectra have exclusively high VLBI-core powers, $>10^{21}$~W~Hz$^{-1}$.

\smallskip
Note. --- Reference:  
\textbf{\citetalias{Herrnstein_etal.1997}}: \citealt{Herrnstein_etal.1997};
\textbf{\citetalias{Cecil_etal.2000}}: \citealt{Cecil_etal.2000};
\textbf{\citetalias{Nagar_etal.2002}}: \citealt{Nagar_etal.2002};
\textbf{\citetalias{Falcke_etal.2000}}: \citealt{Falcke_etal.2000};
\textbf{\citetalias{Ulvestad_Ho2001a}}: \citealt{Ulvestad_Ho2001a};
\textbf{\citetalias{Anderson_etal.2004}}: \citealt{Anderson_etal.2004};
\textbf{\citetalias{Bietenholz_etal.2000}}: \citealt{Bietenholz_etal.2000};
\textbf{\citetalias{Bartel_Bietenholz2000}}: \citealt{Bartel_Bietenholz2000};
\textbf{\citetalias{Ly_etal.2004}}: \citealt{Ly_etal.2004};
\textbf{\citetalias{Giovannini_etal.2001}}: \citealt{Giovannini_etal.2001};
\textbf{\citetalias{Doi_etal.2005}}: \citealt{Doi_etal.2005};
\textbf{\citetalias{Fomalont_etal.2000}}: \citealt{Fomalont_etal.2000};
\textbf{\citetalias{Reid_etal.1989}}: \citealt{Reid_etal.1989};
\textbf{\citetalias{Junor_Biretta1995}}: \citealt{Junor_Biretta1995};
\textbf{\citetalias{vanAlbada_vanderHulst1982}}: \citealt{vanAlbada_vanderHulst1982};
\textbf{\citetalias{Condon_Broderick1988}}: \citealt{Condon_Broderick1988};
\textbf{\citetalias{Ho_Ulvestad2001}}: \citealt{Ho_Ulvestad2001};
\textbf{\citetalias{Condon1987}}: \citealt{Condon1987};
\textbf{\citetalias{Kaufman_etal.1996}}: \citealt{Kaufman_etal.1996};
\textbf{\citetalias{Hummel_etal.1987}}: \citealt{Hummel_etal.1987};
\textbf{\citetalias{Wrobel_Heeschen1984}}: \citealt{Wrobel_Heeschen1984};
\textbf{\citetalias{Laing_Bridle1987}}: \citealt{Laing_Bridle1987};
\textbf{\citetalias{Owen_etal.1989}}: \citealt{Owen_etal.1989}
\end{table}

\subsection{Jet-disk symbioses in LLAGNs}\label{section:individual}

We have suggested in the previous section that jets and the ADAF seem to coexist in LLAGNs.  However, only an unresolved structure has been found in milli-arcsec images for most of the VLBI-detected LLAGNs \citep{Falcke_etal.2000,Ulvestad_Ho2001a,Nagar_etal.2002, Anderson_etal.2004}.  On the other hand, pc/sub-pc-scale jet structures have been found in several LLAGNs.  We find from the core power ranking in Table~\ref{table4} that the presence of VLBI-detectable pc-scale jets depends on the core radio powers.  The pc-scale jets have been found exclusively in high-core-power sources.  In low-core-power sources, no pc-scale jet has been detected in VLBI images even with moderately high dynamic ranges of $\sim100$ \citep{Anderson_etal.2004}.  Cannot the low-core-power sources generate high-brightness jets?

The cases of NGC~3031 and NGC~4258 are very suggestive.  Sub-pc-scale jets have been revealed in their nuclei \citep[e.g.,][]{Herrnstein_etal.1997,Bietenholz_etal.2000}, nevertheless, they are low-core-power LLAGNs.  Note that the two LLAGNs are the nearest two sources in the VLBI-detected sample (Table~\ref{table3}).  We speculate that all low-core-power sources also have sub-pc-scale or smaller jets, and that such structures would be revealed if they were at less than several~Mpc like NGC~3031 and NGC~4258.  The detectability of small-scale jets might merely depend on distance.  If that is true, the 5-GHz VLBI-core power should contain jet contributions significantly.

Additionally, NGC~3031 and NGC~4258 are very suggestive cases also about coexistence with the ADAF disc.  Although they show clear evidence of jets, their high-frequency spectra are highly inverted, $\alpha^\mathrm{15GHz}_\mathrm{96GHz}=0.46\pm0.09$ and $\alpha^\mathrm{15GHz}_\mathrm{96GHz}=0.70\pm0.21$, respectively.  These are the highest two spectral indices in the VLBI-detected LLAGNs with negligible diffuse contaminations.  If their millimetre emissions originate in the ADAF discs, the two LLAGNs would be examples for the coexistence of the jets and ADAF.

We mention apparent exceptions.  NGC~266 and NGC~3147 show steep spectra, but have no detectable jet in spite of their high core powers~(Table~\ref{table4}).  Since the two LLAGNs are the most two distant LLAGNs in our sample, it could be somewhat hard to reveal their putative small-scale jet structures.  Their high radio powers might be also responsible for the dominance of such small-scale jets.  NGC~5866 show a definitely steep spectrum, in spite of a low core power.  Its nuclear activity type, `transition', might be related to the jet domination; another transition object, NGC~4552, also shows a definitely steep spectrum.  However, our limited sample of transition objects cannot allow us to discuss it in detail.

\subsection{Conflict with the unified scheme?}

There appears a clear difference in the spectral properties between type~1s (S1.5, S1.9, and L1.9) and type~2s (S2, L2, and T2), as seen in Table~\ref{table3}.  Most of the type~1s show inverted spectra, while most of type~2s show steep spectra.  We suggest this difference conflicts with the unified scheme \citep{Antonucci1993}.  In this scheme, many observed properties are determined basically only by a viewing angle with respect to the obscuring dusty torus surrounding the nucleus.  Type-1 Seyferts, with broad permitted optical emission lines, are those in which the torus is viewed closer to face-on orientation.  Type-2 Seyferts lack broad permitted lines, except in polarized flux, and thought to be seen with the line of sight passing through the torus in an orientation closer to edge-on.  This classification has been expediently used in LINERs and transitions (e.g., \citealt{Ho_etal.1997b}).  Since the radio emission is transparent through the dusty torus, luminosities and spectral indices of type~1s and 2s should look just the same to the observer unless affected by relativistic beaming.  However, radio powers of type~2s tend to be significantly larger than those of type 1s in our sample.  It seems that the difference is due to larger contribution of jets in type~2s, nevertheless, which would be less affected by relativistic beaming than type~1s.

There are also the other reports about differences between type~1s and type~2s in LLAGNs.  \citet{Ulvestad_Ho2001b} found that low-luminosity type-1 Seyferts rather show somewhat stronger centimetre radio emission than low-luminosity type-2 Seyferts.  \citet{Ho_etal.2001} have reported that type-2 LLAGNs are underluminous in an X-ray band compared to type-1 ones with the same H$\alpha$ luminosity.  There are `intrinsically true type-2 Seyferts (without broad lines even in polarized flux)' especially in the low-luminosity range (\citealt{Tran2003,Laor2003}, and reference therein).  Thus, the discrepancy in the unified scheme is commonly seen in LLAGNs.

We suggest a possible idea for the discrepancy with the unified scheme in terms of radio properties of LLAGNs.  If outflows carry accreting mass away and decrease a density at the innermost disc, then the spectral-peak frequency could decrease to several tens of GHz \citep{diMatteo_etal.2000}.  Although such a central engine could generate strong jets, neither high-frequency inverted spectra nor a lot of up-scattered photons that will ionize broad-line clouds can be produced.  That is, LLAGNs with strong jets would tend to have weak broad-line emission, which could not have been detected in a limited sensitivity of the optical survey, then they could be classified into type 2s, even if the obscuring torus is in a face-on orientation.

\subsection{Implication for the ADAF}

We have found evidence for inverted high-frequency radio spectra in a number of LLAGNs.  Such a high-frequency property is potentially the long-sought signatures of the ADAF.  It is an apparent observational confirmation for the submillimetre bump theoretically-predicted in the ADAF model.  In past observations, the radio spectra of LLAGNs had been investigated exclusively at centimetre bands where high sensitivities are available.  \citet{Nagar_etal.2001} reported the radio spectra of 16 LLAGNs that had been observed with the VLA at 5--15~GHz; \citet{Ulvestad_Ho2001a} reported the radio spectra of three LLAGNs that had been observed with the VLBA at 1.7--8.4~GHz; \citet{Anderson_etal.2004} reported the radio spectra measured with the VLBA for three LLAGNs at 2.3--15~GHz and three LLAGNs at 1.7--43~GHz.  The frequency range of submillimetre-bump hunting has greatly expanded by our high-frequency radio survey.  Our findings will encourage further LLAGN surveys at higher-frequency ranges with, e.g., the coming Atacama Large Millimetre and submillimetre Array (ALMA), to prove the ADAF.

The existence of the ADAF will give radio astronomers a chance to make the direct imaging for accretion discs.  Their area of study had been limited to the nonthermal phenomena in relativistic jets of AGNs, although optical and X-ray astronomers have observed the direct fluxes from the accretion discs as unresolved sources.  Radio astronomers have VLBI techniques, which provide the greatest spatial resolutions all over the electromagnetic wavelengths.  A high-frequency VLBI instrument will report the first image of morphology of accretion discs in micro-arcsec resolutions.   The VSOP-2, a next generation space VLBI mission currently being planned, will provide a 38~micro-arcsec resolution at 43~GHz and a brightness-temperature sensitivity of 10$^8$~K \citep{Hirabayashi_etal.2002}.  Because an electron temperature of 10$^{9-9.5}$~K at the region within several tens Schwarzschild radius is predicted by the ADAF model, the VSOP-2 can reveal disc structures of more than a dozen nearby LLAGNs.  If the accretion disc were an optically-thick standard disc, it will have never been detected with the VLBI because of low brightness temperature.  

From our results, both the jets and disc seem to coexist in the core of LLAGNs, and their flux ratios are different from object to object.  It suggests that relative radiative efficiencies between the jet and disc would determined by some kind of physical parameter(s) in the central engine.  From the theoretical point of view \citep{Narayan_etal.2000,Blandford_Begelman1999}, it has been suggested that radiatively inefficient flows of low accretion rate tend to stimulate strong convective instabilities and powerful outflows.  The central engines of LLAGNs probably can produce jets.  We will discuss the physical parameters that control jet productions in our LLAGN sample in a separate paper (Doi et~al. in~prep).

\section{Summary}\label{section:summary}
Our millimetre survey and analyses of submillimetre archival data for the 20 VLBI-detected LLAGNs reveal high-frequency radio spectra of LLAGNs with compact cores as follows:
\begin{description}
\item[(i)] At least half of the sources show inverted spectra between 15 and 96~GHz.  We used the published data at 15~GHz with the VLA in a 0.15-arcsec resolution and our measurements at 96~GHz with the NMA in a 7-arcsec resolution.
\item[(ii)] Submillimetre fluxes measured with the SCUBA on the JCMT may be from the dust associated with host galaxies.  These data have been useful to estimate dust contamination in the NMA measurements.
\item[(iii)] We confirm little diffuse contamination in the NMA measurements from dust, free--free, and extended nonthermal emissions for LLAGNs showing inverted spectra.  Therefore, the inverted spectra are not caused by resolution effect, but intrinsic properties of the cores.
\item[(iv)] In our sample of 20 LLAGNs, radio cores with inverted high frequency radio spectra reside in at least ten LLAGNs.
\item[(v)] We find an inverse correlation between the high-frequency radio spectrum and the low-frequency radio core power.  Such a property can be explained by an idea that flux ratios between the jet showing steep spectrum and the ADAF disc showing inverted spectrum are different from LLAGN to LLAGN.  The disc components could be seen only if the jets are faint.

\end{description}
Consequently, we have found evidence for inverted high-frequency radio spectra in a number of LLAGNs.  Such a high-frequency property is potentially the long-sought signatures of the ADAF.

\section*{Acknowledgments}
We acknowledge S.~Okumura for helpful advices in observations with the NMA at the Nobeyama Radio Observatory.  The Nobeyama Radio Observatory (NRO) is a branch of the National Astronomical Observatory of Japan (NAOJ), which belongs to the National Institutes of Natural Sciences (NINS).  We have made extensive use of the NVSS and FIRST data from the National Radio Astronomy Observatory that is a facility of the National Science Foundation operated under cooperative agreement by Associated Universities, Inc.  We have also made extensive use of JCMT archival data from Canadian Astronomy Data Centre, which is operated by the Dominion Astrophysical Observatory for the National Research Council of Canada's Herzberg Institute of Astrophysics.

\label{lastpage}

\end{document}